\RequirePackage{fix-cm}

\documentclass[isre,nonblindrev]{informs3}

\DoubleSpacedXI 

\usepackage{graphicx}
\usepackage{float,graphicx}
\usepackage[graphicx]{realboxes}
\usepackage{caption}
\usepackage{booktabs}
\usepackage{float}
\usepackage{subfig}
\usepackage{endnotes}
\usepackage[section]{placeins}
\usepackage{url}
\usepackage{hyperref}
\usepackage{lmodern}
\providecommand{\tightlist}{%
  \setlength{\itemsep}{0pt}\setlength{\parskip}{0pt}}

\usepackage{natbib}
 \bibpunct[, ]{(}{)}{,}{a}{}{,}%
 \def\bibfont{\small}%
\TheoremsNumberedThrough     

\EquationsNumberedThrough    

\usepackage{enumitem}
\usepackage{amsmath}

\begin{document}


\RUNAUTHOR{Li and Sundararajan}

\RUNTITLE{The Rise of Recommerce}


\TITLE{The Rise of Recommerce: Ownership and Sustainability with Overlapping Generations}

\author{Rubing Li}

\ARTICLEAUTHORS{
\AUTHOR{Rubing Li}
\AFF{New York University, \EMAIL{rl4229@stern.nyu.edu}}
\AUTHOR{Arun Sundararajan}
\AFF{New York University, \EMAIL{digitalarun@nyu.edu}}
} 

\ABSTRACT{
The emergence of the branded recommerce channel — digitally enabled and branded marketplaces that facilitate purchasing pre-owned items directly from a manufacturer’s e-commerce site — leads to new variants of classic IS and economic questions relating to secondary markets. Such branded recommerce is increasingly platform-enabled, creating opportunities for greater sustainability and stronger brand experience control but posing a greater risk of cannibalization of the sales of new items. We model the effects that the sales of pre-owned items have on market segmentation and product durability choices for a monopolist facing heterogeneous customers, contrasting outcomes when the trade of pre-owned goods takes place through a third-party marketplace with outcomes under branded recommerce. We show that the direct revenue benefits of branded recommerce are not their primary source of value to the monopolist, and rather, there are three indirect effects that alter profits and sustainability. Product  durability increases, a seller finds it optimal to forgo marketplace fees altogether, and there are greater seller incentives to lower the quality uncertainty associated with pre-owned items. We establish these results for a simple two-period model as well as developing a new infinite horizon model with overlapping generations. Our paper sheds new insight into this emerging digital channel phenomenon, underscoring the importance of recommerce platforms in aligning seller profits with sustainability goals.
}



\maketitle


\newpage

\section{Introduction}

\begin{figure}[H]%
    \centering
    {{\includegraphics[width=7.5cm, height=6.3cm]{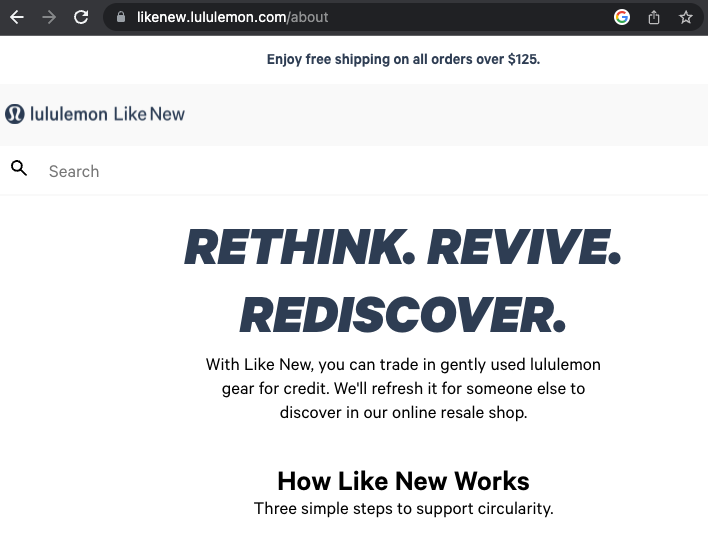} }}%
    \qquad
    {{\includegraphics[width=7.5cm,height=6.3cm]{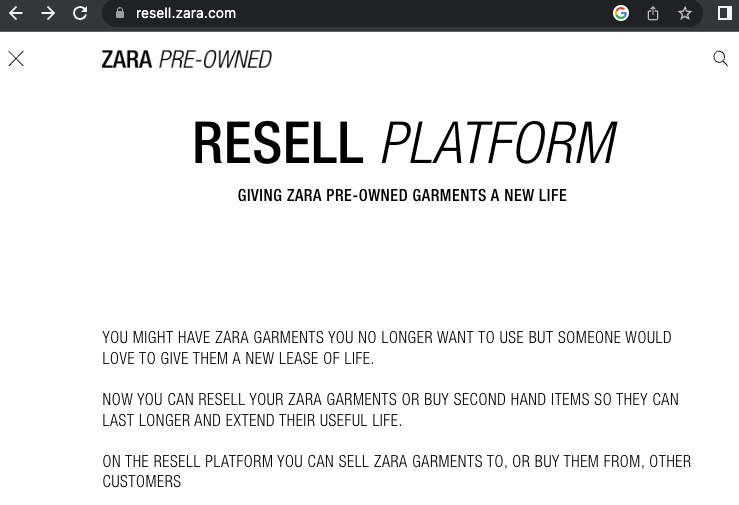} }}%
    \caption
    {Examples of two “branded recommerce” sites. (The Lululemon Like New site is powered by the Trove recommerce platform; the Zara Pre-Owned site is powered by the Recurate recommerce platform.) }%
    \label{fig:recommerce_examples}%
\end{figure}

The market for pre-owned items has seen a significant growth and acceleration of digital innovation over the last few years. Third-party digital markets such as Depop, Poshmark, StockX, and The RealReal have contributed to this recent explosion of “recommerce” — the sale or trading of pre-owned items. In parallel, new digital platforms like Trove and Recurate have made it possible for companies ranging from Canada Goose and Patagonia to Zara and lululemon to set up their own branded marketplaces that facilitate purchasing pre-owned items directly from the brand, a practice we refer to as \emph{branded recommerce}. Figure \ref{fig:recommerce_examples} presents examples of two branded recommerce sites.

Sales of pre-owned apparel, over \$100 billion globally in 2022, are projected to grow at five times the rate of overall commerce, doubling to \$218 billion by 2026 and making up 23\% of apparel retail sales (Thredup, 2022). 

The emergence of platform-based branded recommerce leads to new variants of classic economics and IS questions relating to secondary markets. For example, manufacturers who see branded recommerce as an opportunity to shape the initial brand experience of customers who begin their brand relationship by buying a pre-owned item have to weigh these benefits against the possible increased threat of cannibalization of the sales of new items. A manufacturer who may not have previously encouraged the purchasing of pre-owned items may find their economic incentives are altered by the prospect of value capture from these transactions under branded recommerce, or may see branded recommerce as an “entry deterring” move against specialty third-party marketplaces for pre-owned items. The ability to capture value from recommerce may increase the incentives of a manufacturer to create more durable products better aligned with sustainability goals.

In this paper, our focus is on the interplay of some of these factors with a brand's choice of product \textit{durability}. Intuitively, a more durable product can be used (and reused) more. A higher potential level of reusability can be thought of as being more sustaiable. As an expert on sustainable fashion recently observed in a Bloomberg article (\cite{woody2024}): 

\textit{“Durability and reparability affect environmental impact,” says Veronica Bates Kassatly, a London-based fashion industry analyst focused on sustainability. She cites the example of wearing a shirt 500 times. If a cheaply made shirt is replaced after being worn 10 times, 50 shirts will need to be produced and discarded. But if one better-made and easily mended shirt can be worn 500 times, the environmental cost is significantly lower, even if making the garment uses more water and energy upfront.}

A tradeoff for a brand considering this kind of move towards sustainability stems from the threat of cannibalizing the sales of new products; products that degrade faster help drive more frequent replacement or sales of new products. Perhaps the emergence of a channel that allows a brand to internalize some of the value created by resale alters this tradeoff. 

We develop an economic model of recommerce that sheds light on some of these issues. We begin by examining the effects that the sales of pre-owned items have on market segmentation and product durability choices in a two-period model, contrasting outcomes when the trade of pre-owned goods takes place through a third-party marketplace with outcomes under branded recommerce. Our model includes stylized measures that capture third-party market power (the extent to which value from pre-owned goods trade is captured by the marketplace owner) and quality uncertainty (the gap between the valuation of a pre-owned item by the original owner who knows its actual quality and condition, and a potential buyer who does not). We then extend this analysis to a more general infinte-horizon model in which there are overlapping generations of customers.

Our results show that under fairly general assumptions, adopting branded recommerce is profitable for a monopolist who faces a third-party secondary marketplace. Furthermore, the monopolist’s profit-maximizing durability choices are more aligned with sustainability goals under branded recommerce. We show how this “sustainability bump” from branded recommerce is higher when third-party markets are more powerful, and when pre-owned product quality uncertainty is higher, even though the move to branded recommerce by itself may not resolve any of the underlying quality uncertainty. Counterintuitively, the direct revenue potential of branded recommerce is not its equilibrium source of value. Rather, we demonstrate a set of indirect economic effects which benefit manufacturers and increase sustainability. Product durability increases, there are greater seller incentives to invest in lowering the quality uncertainty associated with pre-owned items, and it is actually optimal for manufacturers to forgo marketplace fees altogether.

\section{Literature Review}

At a time when planned obsolescence is both widespread and lowering product durability is under fire as being environmentally irresponsible, our results suggest that branded recommerce could offer an additional economic lever to sellers balancing profit maximization with sustainability considerations. We build on the theory of the optimal pricing of a product line (\cite{mussa_monopoly_1978}, \cite{maskin_monopoly_1984}, \cite{Moorthy_segmentation_1992}, \cite{Sundararajan_2004}, \cite{Bhargava_Choudhary_2008}), \cite{Mantena_and_saha_2012}, \cite{Pang_and_Etzion_2012}). For example, \cite{waldman_durable_1996} shows that a monopoly seller has an incentive to lower product durability because at higher levels of durability, the ensuing higher quality of the pre-owned item in future periods limits pricing power. We follow \cite{waldman_durable_1996} and \cite{hendel_interfering_1999} in modeling durability as determining the quality gap between new and pre-owned goods, in contrast with \cite{johnson_secondary_2011} and others who treat new and pre-owned goods as having equal quality and model durability as the survival probability of a product. 

Following the literature in economics and IS that has shown the economic importance of product uncertainty (\cite{Pavlou_et_al_2007}, \cite{Dimoka_et_al_2012}, \cite{Ghose_2009}, \cite{Hong_Pavlou_2014}) and transaction cost (\cite{Bakos_1997}), we explicitly incorporate simple representations of these two factors into our model. A more detailed model of how the nature of secondary market frictions affect market outcomes (\cite{Kannan_et_al_2023}) may lead to further insight, and this represents a direction for future researchers.  

Our paper is also related to the literature on the incentives of monopoly sellers of durable goods that stems from the celebrated Coase conjecture (\cite{coase_durability_1972}) — that absent commitment devices, the price charged by a durable goods monopolist is driven down to marginal cost by a time inconsistency problem, thus leading to monopoly choices of durability that are lower than the socially optimal levels, and, as shown by earlier work (\cite{LS_1969}, \cite{schmalensee_regulation_1970}), inducing a monopolist to pursue a strategy of planned obsolescence. A series of papers that followed have either reinforced and refuted the Coase conjecture. For example, in a series of papers, \cite{swan_durability_1970}, \cite{swan_durability_1971}, \cite{swan_optimum_1972} and \cite{sieper_monopoly_1973} assert using a model that endogenizes the monopolist’s choice of durability that this will actually be at the socially optimal level. In contrast, \cite{bulow_durable-goods_1982}, \cite{bulow_economic_1986} reaffirms the Coase results, strengthening the theoretical case for leasing and planned obsolescence. While we do not aim to exhaustively summarize this literature, to the best of our knowledge, ours is the first paper that connects the \emph{ownership} of the secondary market to this discussion on optimal durability. 

Our model with overlapping generations of consumers aims to capture the effect of evolving consumer preferences that \cite{johnson_secondary_2011} and others like \cite{chen2013} and \cite{fraiberger_peer--peer_2015} models as idiosyncratic preference shocks. Our approach follows a standard infinite horizon setting  in which consumers live for two periods (see, for example, \cite{villas-boas_price_2004}).  Rather than incorporating  explicitly changing preferences like some of the prior work on secondary markets, in our model, changes in the “age” of a consumer alter their preferences because their horizon is different (two periods versus one). 

Fundamentally, our paper addresses a variant of the central question in \cite{chen2013}  — whether secondary markets help or hurt producers — albeit with a focus specifically on the marketplace ownership choice (recommerce versus third-party). While simpler, our model captures elements of all three key effects of this literature — the substitution effect (pre-owned items may cannibalize new product sales), the allocative effect (sellers indirectly earn higher rents from higher valuation and forward-looking customers if these customers can sell their products in the future), and aspects of the time consistency effect (this increase in willingness to pay of the higher valuation consumers is affected by their future price expectations, which in turn are shaped by the seller’s output choices, and the seller does not have the ability to commit to future prices).

More saliently, our paper adds to an extensive literature in economics, marketing and information systems on secondary markets (see, for example, \cite{Bapna_et_al_2004}, \cite{Ghose_et_al_2006}, \cite{Ghose_2009}, \cite{rao_understanding_2009}, \cite{gavazza_quantitative_2014}, \cite{chen2013}, \cite{rao_online_2015}, \cite{Pilehvar_et_al_2017}, \cite{ishihara_dynamic_2019}). We also add to the digital platform and sharing economy literature (\cite{Bardhi_et_al_2012}, \cite{cramer_et_al_2016}, \cite{Bapna_et_al_2016}, \cite{Sundararajan_2016}, \cite{Zervas_et_al_2017}, \cite{Li_Wang_2021}). Over the last decade, while true sharing economy platforms that facilitate product sharing and reuse have promised to lead to more sustainable and circular usage, behavior changes that lead to widespread durable goods renting and sharing has been limited outside of the automobile sector. Meanwhile, the trade of pre-owned items has exploded, leading us to believe that perhaps the real impact of platforms on sustainable consumption will come not from sharing but from the digital trade and consumption of pre-owned items. Our efforts to provide a more robust theoretical basis for the ownership structure and economic factors that aid or abet this evolution is aligned with this belief. 
\bigskip

\section{Overview of Model}
\bigskip
\subsection{Firm and customers}

We model a monopoly seller whose products last 2 periods. We begin with a two-period world, and then extend it to an infinite period model with overlapping generations. In the first period, the seller chooses how much durability $D$ to build into its products. This choice is fixed across both periods. \footnote{We have also analyzed the case where the firm makes a separate durability choice in each period. It is of comparable analytical complexity. Since there is no “next period” in the second period, this artificially distorts durability downwards. Further, in practice, it may not be realistic to imagine a firm continuously varying its durability, either for technological reasons, or because of brand perception concerns.} The firm bears a constant marginal cost of production $c(D)$ and no fixed costs. We assume that the function $c(D)$ is positive, increasing and strictly convex for all $D>0$, and thus $c'(0)=0, c'(D)>0, c''(D)>0$ for all $D>0$. We also assume that $c(0)=0$.

New products are of a fixed quality level that is normalized to 1. In the second period, units that are 1 period old are of quality $s(D)$, where $s(0)=0, s(D)<1$ for all $D, s'(D)>0, s''(D)<0$ for all $D \ge 0$. Put simply, the more durable a product, the higher its value as a pre-owned item. 

Customers are heterogeneous, indexed by type $j=L, H$. Each consumer lives 2 periods. There are $n_j$ identical customers of type $\mathrm{j}$. To ensure that the trade of pre-owned goods actually occurs in what follows, we assume that $n_L>n_H$. In each period, type-H customers derive a gross benefit $v_H s$ from consumption of one unit of good of quality $s$, while the corresponding gross benefit derived by the low valuation customers is $v_L s$, $v_H>v_L>0$. Customers use either zero units or one unit of the seller's output in each period. The firm and customers discount the future using the discount factor $\delta, 0<\delta<1$.

\subsection{Interaction Between the Firm and its Customers
}

We assume that the firm cannot explicitly distinguish between consumer types.\footnote{Thus, we ignore the customer recognition information the firm may have in the second period from the choices consumers make in the first period (\cite{villas-boas_price_2004}).} The firm offers one product in each period. Consumers in the second period may also have the option of buying a pre-owned item. Thus, in the second period, the firm faces a design problem analogous to one faced by a multiproduct monopolist, but where the quality level of the inferior version in period 2 is determined by the durability choice made in period 1 and whether the secondary market is run by a third party (more on this later).

The sequence of interaction between the firm and its customers is as follows:

\begin{itemize}
\tightlist
    \item Period 1: The firm first announces a durability choice \(D\), and the price for a new unit of output \(p_{1}^{N}\). All consumers then simultaneously decide what, if anything, to purchase from the firm.
    \item Period 2: First, the firm announces the price \(p_{2}^{N}\) for a new unit of output while maintaining its durability level \(D\). Second, all consumers then decide what, if anything, to purchase directly from the firm. Third, a secondary market opens up where prices on the secondhand marketplace equate supply and demand.
\end{itemize}

Trade in the secondary market is affected by the following economic factors:

\begin{itemize}
\tightlist
    \item \textbf{Quality uncertainty:} Potential buyers of a pre-owned item may not be able to ascertain its true quality prior to purchase. We model this by assuming that in a third-party marketplace, potential buyers of pre-owned goods have a willingness to pay that is deflated by a factor \(\alpha, 0<\alpha<1\), relative to existing owners of these pre-owned goods who know the actual quality. We later assume that when the seller operates the secondary market (the "recommerce case"), this quality uncertainty may change.
    \item \textbf{Marketplace transaction costs:} Sellers of used goods pay a fraction $\beta$, $0<\beta<1$, of the transaction price as a commission. This revenue is received by whoever operates the secondhand marketplace. 
\end{itemize}

The firm chooses durability and prices to maximize total anticipated profits in all remaining periods. The consumers have the following choices. 

\begin{itemize}
\tightlist
    \item In period 2, a consumer who did not purchase a product in the first period chooses between purchasing a new product, purchasing a pre-owned product (if such products are offered) or not participating. A consumer who purchased a product in the first period chooses whether or not to purchase a new product, and whether or not to sell their pre-owned product in the secondary market. 
    \item In period 1, a consumer chooses whether or not to purchase a new product. 

\end{itemize}

\section{A two-period model}

We analyze the two-period model in some detail in what follows, since many of the expressions and results from this analysis carry over to the model with overlapping generations. The following situation —which we refer to as the \textit{active pre-owned marketplace} case — is the one of interest in our analysis, since it is what leads to activity in the marketplace for pre-owned items: 
\begin{itemize}
\tightlist
    \item Period 1: Consumers of type-H purchase a new product from the firm. 
    \item Period 2: Consumers of type-H purchase a new product from the firm, selling their pre-owned products in the secondhand marketplace, and consumers of type-L purchase a pre-owned product. 
\end{itemize}

It is straightforward to show that this case leads to higher firm profits than any case under which the firm's choices lead to type-H consumers wanting to keep their goods in the second period
rather than selling them in the secondhand marketplace. This is because the profits from the latter, \(n_{H}\left[v_{H}(1+\delta s(D))-c(D)\right]\), are strictly less than the profits \((1+\delta) n_{H} v_{H}\) that the firm could earn from offering a product of durability zero in each period priced at \(v_{H}\). There is yet another possibility - that the firm chooses a strategy that induces type-L consumers to purchase new goods in one or both periods. However, under any strategy of this kind, there will be no trade in the secondhand marketplace: if type-L consumers purchase a new good in period 1, there is no difference between the pre-owned goods being offered for sale and the ones they hold in period 2 as a consequence of their period 1 purchase, and if type-L consumers do not purchase pre-owned goods in period 2 , there are no potential buyers for any pre-owned goods being offered for sale in period 2. We examine this in greater detail in our overlapping generations model in Section 4 .

\subsection{Socially optimal product choices}

We benchmark the socially optimal level of durability under this active pre-owned marketplace setting — where type-H customers consume a new good in each period, and where type-L customers do not participate in the market in the first period, while buying a pre-owned good in the second period.\footnote{This may seem like a somewhat unusual socially optimal benchmark. However, it is the one that is meaningful in our context. In contrast, if we consider the situation under which both consumer types consumed a new good in each period, durability is irrelevant, and the optimal level of durability is zero. To motivate this benchmark, imagine that the seller can identify types, creates a single period rental contract for the new good offered to type-H in each period, and in the second period, creates a single-period rental contract that offers the pre-owned goods returned by the type-H consumer at the end of period 1 to type-L customers in period 2.  This situation yields what is akin to the “first best” durability choice.}

Under this benchmark setting, viewing things from the first period, suppose the firm chooses a
durability level \(D\). The gross surplus from the consumption of the type-H customers is \(n_{H} v_{H}\) in
each period. The total cost borne by the seller from this consumption is \(c(D)\) in each period.
Since \(n_{L}>n_{H}\), only a fraction \(n_{H} / n_{L}\) of type-L customers will have a pre-owned good available
for their consumption in period 2. The gross surplus generated from this consumption is
\(n_{H} v_{L} s(D)\). Thus, the total surplus from this consumption and production activity, viewing things
from the first period is:
\begin{equation}
\label{eqn:total_surplus}
(1+\delta) n_{H} v_{H}+\delta n_{H} v_{L} s(D)-(1+\delta) n_{H} c(D).
\end{equation}
Maximizing (\ref{eqn:total_surplus}) with respect to $D$ yields the following expression for the socially optimal durability level $D^{* *}$:
\begin{equation}
\label{eqn:social_optimal_durability}
c^{\prime}\left(D^{* *}\right)=\frac{\delta}{1+\delta} v_{L} s^{\prime}(D^{* *}).
\end{equation}

\subsection{Prices under the active digital pre-owned marketplace case}

Let \(p_{2}^{n}\) denote the price for a new good in the second period, and \(p_{2}^{u}\) denote the price for a pre-owned good in the second period.  One can formulate the seller’s problem in period 2 in terms of incentive-compatibility (IC) and individual rationality (IR) constraints that induce each type to choose the product “designed” for them. (This is not essential, but we present it this way so that the parallel with the standard price screening problem is clearer to those readers familiar with that literature.) 
\begin{itemize}
\tightlist
    \item In period 2, type-H customers choose to sell their pre-owned product and buy a new product rather than holding on to their pre-owned product (akin to an IC condition for type-H)
    \begin{equation}
    \label{eqn:IC_H}
v_{H}-p_{2}^{n}+(1-\beta) p_{2}^{u} \geq v_{H} s(D).
\end{equation}
    \item In period 2, type-L customers choose to buy a pre-owned product rather than a new product (akin to an IC condition for type-L)
    \begin{equation}
    \label{eqn:IC}
\alpha v_{L} s(D)-p_{2}^{u} \geq v_{L}-p_{2}^{n}.
\end{equation}
    \item The surplus of each type in period 2 is non-negative (akin to the IR conditions)
\begin{equation}
v_{H}-p_{2}^{n} \geq 0,
\end{equation}
\begin{equation}
\alpha v_{L} s(D)-p_{2}^{u} \geq 0.
\end{equation}
\end{itemize}

It is straightforward to see that the IR condition for type-L will bind, and thus:
\begin{equation}
\label{eqn:p2u}
p_{2}^{u}=\alpha v_{L} s(D).
\end{equation}
Next, we can use (\ref{eqn:p2u}) to rewrite equation (\ref{eqn:IC_H}) as
\begin{equation}
v_{H}-p_{2}^{n} \geq v_{H} s(D)-\alpha(1-\beta) v_{L} s(D),
\end{equation}
which yields the maximum price the firm can charge for a new good of durability $D$ in period 2 \footnote{This is akin to the IC condition for type-H binding.}:
\begin{equation}
\label{eqn:p2n}
p_{2}^{n}=\alpha(1-\beta) v_{L} s(D)+v_{H}(1-s(D)).
\end{equation}

\subsection{Durability choices}

Let $D^*$ denote the firm’s profit-maximizing durability choice. We can now characterize the optimal choices of the seller under two regimes: when the secondhand marketplace is run by a third-party (“third-party marketplace”) and when the secondhand marketplace is run by the seller (“branded recommerce”). 

\begin{lemma}(Equilibrium outcomes with a third-party marketplace): 

If \(v_{L}>v_{H} / 2 \alpha(1-\beta)\), then
\begin{enumerate}[label=(\alph*)]
    \item The profit maximizing level of durability $D^*$ satisfies: 
    \begin{equation}
c^{\prime}\left(D^{*}\right)=\frac{\delta}{1+\delta}\left[2 \alpha(1-\beta) v_{L}-v_{H}\right] s\left(D^{*}\right).
\end{equation}
    \item The price of new goods in period 1 is \begin{equation}
p_{1}^{n}=v_{H}+\delta \alpha(1-\beta) v_{L} s(D^{*}).
\end{equation}
 \item The price of new goods in period 2 is  \begin{equation}
p_{2}^{n}=\alpha(1-\beta) v_{L} s(D^{*})+v_{H}(1-s(D^{*})).
\end{equation}
\end{enumerate}
\end{lemma}

Note that if $v_L \leq v_H / 2 \alpha(1-\beta)$, there is no positive value of $\mathrm{D}$ that supports an active pre-owned marketplace, since it is more profitable for the seller to choose a durability level of zero, setting prices $p_1^n=p_2^n=v_H$ and shutting the lower types out of the market.

We now turn to the branded recommerce situation under which the seller operates the pre-owned marketplace. To aid in comparability with the prior case, we retain the possibility of some quality uncertainty remaining even if the seller operates the pre-owned marketplace. We discuss the case of $\alpha = 1 $ as a special case later in the paper.

\begin{lemma}(Equilibrium outcomes with branded recommerce): 

If \(v_{L}>v_{H} / \alpha(2-\beta)\), then
\begin{enumerate}[label=(\alph*)]
    \item The profit maximizing level of durability $D^*$ satisfies: 
    \begin{equation}
c^{\prime}\left(D^{*}\right)=\frac{\delta}{1+\delta}\left[ \alpha(2-\beta) v_{L}-v_{H}\right] s\left(D^{*}\right)
\end{equation}
    \item The price of new goods in period 1 is \begin{equation}
p_{1}^{n}=v_{H}+\delta \alpha(1-\beta) v_{L} s(D^{*})
\end{equation}
 \item The price of new goods in period 2 is \begin{equation}
p_{2}^{n}=\alpha(1-\beta) v_{L} s(D^{*})+v_{H}(1-s(D^{*}))
\end{equation}
\end{enumerate}
\end{lemma}

Akin to the case of a third-party marketplace, if \(v_{L} \leq v_{H} / \alpha(2-\beta)\), then it is more profitable for
the seller to choose a durability level of zero, setting prices \(p_{1}^{n}=p_{2}^{n}=v_{H}\) and shutting the
lower types out of the market. Since our focus is on examining the difference between
third-party and branded recommerce, we do not analyze this case further.

\subsection{Contrasting branded recommerce with third-party pre-owned markets}

We now turn to analyzing how the possibility of branded recommerce changes a firm’s choices. First, we examine how the pre-owned marketplace commissions and quality uncertainty distort the durability choices of the seller, independent of who runs the pre-owned marketplace. Next, we establish how branded recommerce leads to more sustainable choices by inducing an increase in the durability of products (and as a consequence, the quality of pre-owned goods) and separate out three different drivers of this increase.  

\begin{proposition} Independent of whether the seller or a third-party operates the pre-owned marketplace:
\label{proposition_1}
\begin{enumerate}[label=(\alph*)]
    \item As quality uncertainty goes down (and thus, $\alpha$ increases), the optimal level of durability $D^*$ chosen by the seller increases,  and the seller profit increases.
    \item As the commission $\beta$ charged by the operator of the pre-owned marketplace increases, the optimal level of durability $D^*$ chosen by the seller decreases, and the seller profit decreases. 
\end{enumerate}
\end{proposition}

  The proof of the proposition is provided in Appendix. Proposition \ref{proposition_1} indicates that the emergence of robust third-party pre-owned marketplace can have countervailing effects on a seller’s sustainability incentives. On the one hand, the emergence of a pre-owned marketplace where none existed in the past has a strictly positive effect on the durability, raising it above the baseline level of 0 in our model \footnote{One might ask whether we are ignoring, in the absence of a pre-owned marketplace, the possibility of the seller creating a product with positive durability aimed at getting higher type consumers to purchase the product and hold it for 2 periods. This option is strictly dominated by the options of offering a product of durability 0 in both periods, priced at $v_H$.}.  However, an increase in the power of a third-party marketplace operator (leading to an increase in the commission rate ) lowers the seller’s incentives to create a more durable (and thus sustainable) product. 

We now show how these potentially countervailing incentives are all aligned when the firm runs its own pre-owned marketplace.

\begin{proposition}
\label{proposition_2}
    Relative to the case of a third-party pre-owned marketplace, under the case of branded recommerce:
    
    \begin{enumerate}[label=(\alph*)]
    
\item Optimal durability is higher at any third-party commission rate $\beta$ and quality uncertainty $\alpha$.
\item The seller will choose a commission rate $\beta = 0.$ 
\item The incentives of a seller to invest in reducing quality uncertainty $\alpha$ are higher. 

\end{enumerate}

\end{proposition}

Proposition \ref{proposition_2} establishes a main result — that branded recommerce has a \textit{direct effect} of inducing higher durability, or put differently, increases the incentives of a firm that makes something to design it to last longer. It also proves a set of indirect economic effects which benefit manufacturers and increase sustainability. Part (b) shows that it is optimal for manufacturers to forgo marketplace fees altogether. Counterintuitively, therefore, the direct revenue potential of branded recommerce is not its equilibrium source of value. Part (c) shows that there are greater seller incentives to invest in lowering the quality uncertainty associated with pre-owned items. The “branding” of recommerce by itself may reduce quality uncertainty — after all, a customer is buying the pre-owned item directly from the manufacturer rather than from an unknown individual.  However, a key implication of part (a) of the proposition is that the welfare gains of branded recommerce from an increase in durability do not have to assume any change in perceived quality, and the incentive to lower quality uncertainty is a separate positive economic effect.   
In what follows, we examine whether the insights gained from our two-period model carry over to a more general setting.

\section{A model with overlapping generations}

In this section, we develop a model of infinitely many periods in which a fixed number of customers enter in each period, and each generation of customers lives for two periods. We continue to assume that the firm sets a level of durability \(D\) that remains fixed across periods. In each period \(t\), the seller chooses its price \(p^{n}(t)\), bearing a constant marginal cost of production
\(c(D)\) that is positive, increasing and strictly convex for all \(D>0\). Products last two periods. In each period \(t\), new units of output are of a fixed quality level that is normalized to 1 , and pre-owned units (those produced in period t-1) are of quality \(s(D)\), where, as before,
\(s(0)=0, s(D)<1, s^{\prime}(D)>0, s^{\prime \prime}(D)<0\) for all \(D \geq 0\). The price of pre-owned items in period
\(\mathrm{t}\) is \(p^{u}(t)\).

As before, the fraction of type-H (L) customers that enter in each period is \(n_{H}\left(n_{L}\right)\), where
\(n_{H}+n_{L}=1\). Type-H customers derive a gross benefit per period of \(v_{H} s\) from consuming a good of perceived quality \(s\), while the corresponding gross benefit derived by type-L customers is \(v_{L} s\),
and \(v_{H}>v_{L}>0\). In any "current" period, the customers who entered in the previous period are referred to as "age-2" customers, and those customers who enter in the current period are referred to as "age-1" customers. Customers and sellers discount the future at the rate $\delta$.

The payoff-relevant state variable \(x(t)\) for period \(t\) is the fraction of age-2 customers who bought the new product in the previous period \(t-1\). We focus on deterministic consumer and seller choices. As before, the seller cannot observe customer types, which means that if type-L customers purchase in any period, then so do type-H customers. Thus, the possible states are
\(\mathrm{x}(\mathrm{t}) \in\left\{0, n_{H}, 1\right\}\). 

\begin{figure}
    \FIGURE
    {\includegraphics {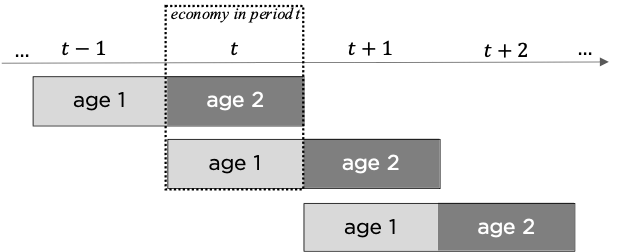}}
    {An Illustration of the overlapping generation model \label{figure_1}}
    {It illustrates the overlapping of generations in each period. In period t, the prior generation that was born in period (t — 1) is of age 2, and the new generation that is born in period t is of age 1. In every period, the seller thus faces both “legacy” customers and “fresh” customers.}
\end{figure}

Following the standard Blackwell/Bellman approach, we look for a stationary policy which maps each state \(\left\{0, n_{H}, 1\right\}\) to a period price \(p^{n}(t)\) for the new good. We begin with the case of \(\mathrm{D}=0\). The state will always be \(\mathrm{x}=0\), and the seller's choices here are therefore to price the product at either \(v_{L}\) or \(v_{H}\). Thus,the optimal steady-state policy will be to price at \(v_{L}\) if \(v_{L}>n_{H} v_{H}\) and at \(v_{H}\) otherwise. Now consider \(\mathrm{D}>0\). In each period apart from the first period, the seller faces a setting somewhat similar to what was faced in period 2 of the two-period model. However, there are, essentially, four types of customers rather than two - age-1 type-L, age-1 type-H, age-2 type-L and age-2 type-H. 

The set of possible choices that each type-age pair has available to them are summarized below.

\begin{table}[H]
\centering
\caption{Set of possible choices when \(x=n_{H}\) \label{tab:Table_1}}
{\begin{tabular}{|c|c|c|}
\hline & \textbf{Type-L} & \textbf{Type-H}\\ \hline \textbf{Age 2 }& \{buy new, buy used, do nothing\} & \{sell used+buy new, buy new, keep used\} \\ \hline \textbf{Age 1} & \{buy new, buy used, do nothing\} & \{buy new, buy used, do nothing\} \\ \hline\end{tabular}
}
\end{table}
\begin{table}[H]
\centering
    \caption{Set of possible choices when \(x=0\)}
    \label{tab:Table_2}
    \begin{tabular}{|c|c|c|}\hline & \textbf{Type-L} & \textbf{Type-H }\\ \hline \textbf{Age 2 } & \{buy new, buy used, do nothing\} & \{buy new, buy used, do nothing\} \\ \hline \textbf{Age 1} & \{buy new, buy used, do nothing\} & \{buy new, buy used, do nothing\} \\ \hline
    \end{tabular}
\end{table}

\begin{table}[H]
    \centering 
    \caption{Set of possible choices when \(x=1\)}
    \label{tab:Table_3}
\begin{tabular}{|c|c|c|}\hline & \textbf{Type-L} & \textbf{Type-H }\\ \hline \textbf{Age 2 } & \{sell used+buy new, buy new, keep used\} & \{sell used+buy new, buy new, keep used\} \\ \hline \textbf{Age 1} & \{sell used+buy new, buy new, keep used\} & \{sell used+buy new, buy new, keep used\} \\ \hline\end{tabular}
\end{table}

We next characterize the properties of an optimal steady-state policy. Under such a policy, the seller prices the new goods at the same level in each period, and this price leads to choices by the customers that maintain the same state. 

\begin{lemma}The only state that can be part of a steady-state policy is one in which all customers with higher willingness to pay (Type-H) purchase in each period, that is, \(x=n_{H}\)
\end{lemma}

We present the proof in the main body of the paper since it is simple and aids in understanding what follows. Clearly, a steady-state in which \(x=0\) is suboptimal for the seller, since it yields no profits. Consider next a steady state in which \(x=1\). Referring to Table \ref{tab:Table_3} , it is clear that both type-L and type-H customers of age-1 must choose to buy new, since the state must remain at \(x=1\). However, if this is in fact a steady-state, then both type-L and type-H customers of age-2 already own a pre-owned item. As a consequence, there are no potential buyers for the pre-owned items in any period. The profit-maximizing price that sustains this steady-state is \(v_{L}(1+\delta s(D))\) which yields a demand level of 1 in each period. Following an argument similar to one made in Section 4, the profit stream from this price is strictly lower than the profit stream from a choice of \(D=0\) and a price \(v_{L}\) which yields a demand level of 2 in each period. This completes the proof.

Next, let us consider the set of consumer actions that are consistent with a steady-state policy. We show in our next result that under any steady-state policy in which there is any kind of pre-owned market activity, customers with lower willingness to pay will always buy pre-owned items while the choices of customers with higher willingness to pay depends on their decision horizon. 

\begin{lemma}
\label{lemma4}
    Under any steady-state policy
\begin{itemize}
    \item All customers with lower willingness-to-pay (type-L customer) will purchase a pre-owned item that lasts one period. 
    \item Customers with a higher willingess-to-pay who have just entered the market (Type-H, age 1) will buy a new product
    \item Customers with a higher willingess-to-pay who were in the market in the previous period (Type-H, age 2) will sell their existing product and buy a new product

\end{itemize}
\end{lemma}

Again, rather than relegating this proof to the Appendix, we present it in the main body of the paper because it aids in understanding the implications of the result. 
 
First, to maintain the steady-state \(x=n_{H}\), it is clear that type-H customers of age-1 must choose to buy new. The choices of type-H customers of age 2 can be partitioned as follows: 

\begin{itemize}
\tightlist
    \item If type-H customers of age-2 discard their existing product and buy new, the steady-state
then involves type-\(L\) customers not participating, and type-H customers consuming their
product for one period. In this case, any durability level \(D>0\) is strictly less profitable for the
seller than the choice of \(D=0\).
\item  If type-H customers of age 2 keep their existing product, then the steady-state again
involves type-L customers not participating, and type-H customers consuming their product
for both periods of their life. The profit-maximizing price that sustains this steady-state is
\(v_{H}(1+\delta s(D))\) which yields a demand level of \(n_{H}\) per period. However, this is strictly less
profitable for the seller than choosing a durability \(\mathrm{D}=0\), pricing at \(v_{H}\), and realizing a
demand of \(2 n_{H}\) (from both all type-H customers).
\item  If type-H customers of age 2 sell their existing product and buy a new product, then one
must characterize the choices of type-L customers (since someone has to buy these
products). Since the product lasts 2 periods, a pre-owned product has a remaining life of 1
period. As a consequence, its value is identical to both age 1 and age 2 customers of
type-L.
\begin{itemize}
\tightlist
    \item The steady-state cannot be maintained if type-L customers of age 1 prefer to buy the new product. 
    \item The steady-state can be maintained if type-L customers of age 2 prefer to buy the new product at whatever price levels the new and pre-owned products are at (since they are no longer in the economy after the current period, their choices do not affect the state in the next period), while type-L customers of age 1 prefer to purchase the pre-owned product. However, since type-L customers of age 1 have an additional period of life and thus have a strictly higher consumption value from the new product, which in turn means that the  set of actions (type-L customers of age 2 buy new, type-L customers of age 1 buy pre-owned) is not individually rational. 
   \item The steady state can also be maintained if type-L customers of both age 1 and age 2
prefer to buy the used product over buying the new product. This is possible if there are
sufficient type-L customers, that is, if \(2 n_{L}>n_{H}\), or if \(n_{H}<2 / 3\)
\end{itemize}

\end{itemize}

This completes the proof. 

The choices implied by Lemma 4 are summarized in Table \ref{tab:Table_4}

\begin{table}[H]
    \centering
    \caption{Choices made by each type under any steady-state policy for which $D > 0$}
    \label{tab:Table_4}
\begin{tabular}{|c|c|c|}\hline & \textbf{Type-L} & \textbf{Type-H } \\ \hline \textbf{Age 2 }& buy used & sell used+buy new \\ \hline \textbf{Age 1} & buy used & buy new \\ \hline\end{tabular}

\end{table}

Next, we characterize the prices \(p^{n}, p^{u}\) that are consistent with these choices. We reintroduce the marketplace commission $\beta$ and quality uncertainty $\alpha$ parameters defined in Section 2. A subset of the conditions that define the consumer choices are summarized below.
\begin{align}
    &\text{[Type-H, Age 2]: }  v_{H}-p^{n}+(1-\beta) p^{u} \geq v_{H} s(D). \label{eqn:H_2}\\
    &\text{[Type-H, Age 1]: }  v_{H}-p^{n}+\delta\left(\max \left[v_{H}-p^{n}+(1-\beta) p^{u}, v_{H} s(D)\right]\right) \geq v_{H} s(D)-p^{u}. 
    \label{eqn:H_1}\\
    &\text{[Type-L, both ages]: }  \alpha v_{L} s(D)-p^{u} \geq 0.
    \label{eqn:L_12}\\
    &\text{[Type-L, Age 1]: }  \alpha v_{L} s(D)-p^{u} \geq v_{L}-p^{n}+\delta\left(\max \left[v_{L}-p^{n}+(1-\beta) p^{u}, v_{L} s(D)\right]\right).
    \label{eqn:L_1}\\
    &\text{[Type-L, Age 2]: }  \alpha v_{L} s(D)-p^{u} \geq v_{L}-p^{n}.
    \label{eqn:L_2}
\end{align}

(\ref{eqn:H_2}) and (\ref{eqn:H_1}) are analogous to the IC conditions for the type-H customers. Since \(v_{H}>v_{L}>0\),
(\ref{eqn:H_2}), (\ref{eqn:H_1}) and (\ref{eqn:L_12}) also imply that type-H consumers of both ages get non-negative surplus from their choice. (\ref{eqn:L_12}) ensures that type-L customers get non-negative surplus, and (\ref{eqn:L_1}) and (\ref{eqn:L_2})
are analogous to their IC conditions. It is straightforward to see that if (\ref{eqn:L_1}) is satisfied, (\ref{eqn:L_2}) can
be ignored. 

Our final lemma establish the prices consistent with the steady-state described in Lemma 4 and summarized in Table 4. 

\begin{lemma}The prices of the new and pre-owned item consistent with a steady-state are identical to those obtained in the two-period model:
\label{Lemma5}
\begin{equation}
p^u=\alpha v_L s(D)
\end{equation}
\begin{equation}
p^n=\alpha(1-\beta) v_L s(D)+v_H(1-s(D))
\end{equation}

\end{lemma}

We are now in a position to contrast the outcomes under branded recommerce with those obtained when there is a third-party marketplace. By Lemma \ref{Lemma5}, under the steady-state policy, the firm’s per-period payoffs with a third-party marketplace are:
\begin{equation}
n_{H}\left[\alpha(1-\beta) v_{L} s(D)+v_{H}(1-s(D))-c(D)\right],
\end{equation}
and the firm's per-period payoffs under branded recommerce after taking quality undertainty and transaction fees into consideration are
\begin{equation}
n_{H}\left[\alpha v_{L} s(D)+v_{H}(1-s(D))-c(D)\right].
\end{equation}

Notice the close parallel between these expressions and those derived in the two-period model. Happily, we can now establish that under the optimal steady-state policy, a set of results in Proposition \ref{proposition_3} that parallel those established for the two-period model continue to hold.  

\begin{proposition}
\label{proposition_3}
     In a model with overlapping generations, under the unique optimal steady-state policy:
    \begin{enumerate}[label=(\alph*)]
    
\item Independent of the ownership structure of the third-party marketplace, as quality uncertainty goes down (and thus, $\alpha$ increases), the optimal level of durability $D^*$ chosen by the seller increases, and the seller profit increases.
\item Independent of the ownership structure of the third-party marketplace, as the commission $\beta$ charged by the operator of the pre-owned marketplace increases, the optimal level of durability $D^*$ chosen by the seller decreases, and the seller profit decreases. 
\item Under branded recommerce: (i) the durability level chosen by the seller is higher, (ii) the seller will choose a commission rate $\beta=0$ , and the incentives of a seller to invest in reducing quality uncertainty are higher. 
\end{enumerate}

\end{proposition}

Proposition 3 establishes that the main results of Proposition 2 -- that choosing \emph{branded recommerce} has the direct effect of inducing the manufacturer of a product to make more durable products -- is not a modeling artifact of a two-period setting, but generalizes to an infinite horizon setting with overlapping generations Broadly, this result underscores what could be a key societal benefit of branded recommerce. Granted, sellers may be drawn to branded recommerce by the prospect of greater control of the customer experience for those buyers who begin their brand interaction by purchasing pre-owned items, or by the promise of profits from secondary market transactions. Our result establishes that there is a path to greater profits from sacrificing the latter, and that in addition to whatever firm-specific benefits that may accrue from branded recommerce, it shifts incentives in a manner that makes it more desirable for a seller to make goods that last longer, rather than resorting to the less sustainable path of more perishable durables in pursuit of repeat sales.

\section{Conclusion}

The promise of digitally-enabled branded recommerce is clear to a brand — running one’s own pre-owned marketplace allows value capture from the trade of one’s used items, signals support for  circular consumption and allows stronger brand experience control for the growing subset of consumers who favor buying pre-owned items. Brands, however, may initially balk at the prospect of placing pre-owned items aside new products on their ecommerce sites, fearing profit reductions from cannibalization. 

Our analysis presents reassuring news: that branded recommerce has a number of positive effects on both durability and profitability. Interestingly, we find that a brand’s focus should not be on the value capture promise of branded recommerce — in our stylized setting, a seller finds it optimal to forgo marketplace fees altogether, suggesting that direct revenue impacts of branded recommerce should not be seen as their primary economic benefit. Rather, we uncover more subtle benefits that stem from the ability to set higher prices, the ability to profitably raise product durability, and greater seller incentives to lower quality uncertainty associated with pre-owned items. 

While formal empirical validation of our theoretical results remains in the future, it has been reassuring to observe anecdotal evidence of a relationship between the adoption of branded recommerce and investments into raising product durability. For example, Patagonia, which pioneered its "Worn Wear" effort to sell pre-owned items in stores in 2013, and was of the early adopters of branded recommerce in 2017, has since made significant efforts to increase the durability and reusability of the products it manufactures (\cite{woody2024}). 

Our future work aims to extend our model in a few directions. We are working on an extension that considers the potential simultaneous presence of  branded recommerce and a third-party marketplace. Our preliminary findings suggest that under certain conditions, the threat of a third-party marketplace can cause a firm to introduce branded recommerce in settings under which it would not have customarily encouraged the trade of pre-owned items -- this move can deter the entry of a third-party marketplace that would have a more adverse effect on seller outcomes. Our model extension in which the two pre-owned marketplace options are differentiated by both quality uncertainty and product availability is in progress, but is one we hope will take us closer to having richer testable results.  

We are also exploring extensions to the baseline model that add additional details that reflect what we have observed in practice. For example, in the implementation of recommerce, there may be differing levels of search frictions (\cite{Gavazza_2021}, \cite{Rhodes_Watanabe_Zhou_2021}) associated with the pre-owned item when compared to new items, depending on whether the pre-owned item is listed alongside the new item (like it is on Amazon), in a separate tab on the brand's primary e-commerce site (like it is on Lululemon), or on an entirely different web site. While we currently model this as being subsumed by the reduction in perceived value, a more detailed model could shed new insight and lead to testable hypotheses, given the variations in these frictions we have observed across recommerce sites. The choices on these front may also have quality signaling implications (\cite{Nian_Sundararajan_2022}, \cite{Ananthakrishnan_et_al_2023}), as firms that highlight their recommerce sites more visibly could be perceived as standing more firmly behind the quality of their durable goods. There are often unexpected and subtle economic effects from changes to market participant visibility (\cite{Bapna_et_al_2016}). Many of these will eventually be empirical questions, and we have partnered with a leading recommerce platform to test some of our current and ongoing theory results. 

While research into digital secondary markets has been widespread in economics, marketing and IS over the last two decades, the phenomenon of branded recommerce is new, and, powered by new recommerce platforms, will account for tens of billions of dollars of ecommerce in the coming years. Branded recommerce is the most significant new digital “channel,” as important today as ecommerce was two decades ago. Our paper is the first we are aware of that sheds new insight into this important new digital development, underscoring its importance in aligning seller profits with sustainability goals.

\ACKNOWLEDGMENT{We thank Anindya Ghose, Silin Huang, Xiao Liu, Vandith Pemuru, Andy Ruben, Joao Sedoc, Zhengyuan Zhou and seminar participants at New York University, the 2024 International Industrial Organization Conference, and the 2023 Marketing Science, ICIS and WISE conferences for helpful comments and discussion. 
}

%

 \begin{APPENDIX}{Technical Details and Proofs}

\textbf{\textit{Proof of Lemma 1}.} 
 Recall that in the active pre-owned marketplace case, type-H customers purchase new goods in period 1 and type-L customers purchase the pre-owned goods from type-H customers via the third-party marketplace in period 2. We derive the optimal level of durability and the optimal prices of new goods in each period under this setting.

Consider any durability choice D. From (\ref{eqn:p2n}), the corresponding price of new goods in the second period is:
\begin{equation}
p_2^n=\alpha(1-\beta) v_L s(D)+v_H(1-s(D)) . \nonumber
\end{equation}
The firm’s profits in second period are therefore
\begin{equation}
n_H\left[\alpha(1-\beta) v_L s(D)+v_H(1-s(D))-c(D)\right] . \nonumber
\end{equation}
If the firm and customers anticipate the second period, since the type-L customers are not active in the first period, the maximum price the firm can charge for new units in the first period is the value realized from the new good in the first period added to the discounted price the consumer will receive from selling the good in the pre-owned marketplace in period 2, or 
\begin{equation}
p_1^n=v_H+\delta \alpha(1-\beta) v_L s(D),\nonumber
\end{equation}
and the firm’s profits from these first period sales are therefore
\begin{equation}
n_H[v_H+\delta \alpha(1-\beta) v_L s(D)],\nonumber
\end{equation}

Thus, the firm’s ex ante total expected profits at the time they choose $D$ is: 

\begin{equation}\tag{A1}
\pi(D)=n_H\left[v_H+\delta \alpha(1-\beta) v_L s(D)-c(D)\right]+\delta n_H\left[\alpha(1-\beta) v_L s(D)+v_H(1-s(D))-c(D)\right]
\label{eqn:A1}
\end{equation}

The firm’s optimal durability choice $D^*$ maximizes (\ref{eqn:A1}). The first-order condition yields:
\begin{equation}\tag{A2}
c^{\prime}\left(D^*\right)=\frac{\delta}{1+\delta}\left[2 \alpha(1-\beta) v_L-v_H\right] s^{\prime}\left(D^*\right)
\label{eqn:A2}
\end{equation}
Now, since $v_L>v_H / 2 \alpha(1-\beta)$, it follows that $2 \alpha(1-\beta) v_L-v_H>0$, and since $c^{\prime}(0)=0$ and $c^{\prime \prime}(D)>0$, there is a positive value of $D^*$ that satisfies (\ref{eqn:A2}).
Comparing \ref{eqn:social_optimal_durability} and (\ref{eqn:A2}), since $c^{\prime \prime}(D)>0$, it follows that $D^*<D^{* *}$ if
$$
\left[2 \alpha(1-\beta) v_L-v_H\right]<v_L
$$
which simplifies to
$$
v_H>v_L[2 \alpha(1-\beta)-1],
$$
which is true for all $0<\alpha<1,0<\beta<1$, since $v_H>v_L$. (Q.E.D.)

\textbf{\textit{Proof of Lemma 2}.} 
 This proof parallels that of Lemma 1. Following the same logic, the first and second period price for new goods are, respectively

\begin{equation}
\begin{aligned}
& p_1^n=v_H+\delta \alpha(1-\beta) v_L s(D) \\
& p_2^n=\alpha(1-\beta) v_L s(D)+v_H(1-s(D))
\end{aligned}\nonumber
\end{equation}

As in Lemma 1, the firm’s first period profits are $n_H\left[v_H+\delta \alpha(1-\beta) v_L s(D)-c(D)\right]$. However, in the second period, as the operator of the pre-owned marketplace, the firm also receives marketplace commissions $n_H \beta p_2{ }^u$. (Note that only a fraction $n_H / n_L$ of type-L customers end up purchasing the pre-owned product owing to this being the total supply.) Recall from (\ref{eqn:p2u}) that $p_2{ }^u=\alpha v_L s(D)$. The firm's second period profits are therefore:
$$
\pi=n_H\left[\alpha(1-\beta) v_L s(D)+v_H(1-s(D))-c(D)\right]+\alpha \beta v_L s(D),
$$
which simplifies to:
$$
\pi=n_H\left[\alpha v_L s(D)+v_H(1-s(D))-c(D)\right] .
$$
Thus, the firm's ex ante expected profits in first period when choosing durability D are
$$
\pi=n_H\left[\alpha(1-\beta) v_L s(D)+v_H(1-s(D))-c(D)\right]+\alpha \beta v_L s(D),
$$
which simplifies to:
$$
\pi=n_H\left[\alpha v_L s(D)+v_H(1-s(D))-c(D)\right] .
$$

Thus, the firm’s ex ante expected profits in first period when choosing durability $D$ are 

\begin{equation}\tag{A3}
\pi(D)=n_H\left[v_H+\delta \alpha(1-\beta) s(D) v_L-c(D)\right]+\delta n_H\left[\alpha v_L s(D)+v_H(1-s(D))-c(D)\right] .
\label{eqn:A3}
\end{equation}

The firm’s optimal durability choice $D^*$ therefore maximizes (\ref{eqn:A3}). The first-order condition yields:

\begin{equation}\tag{A4}
c^{\prime}\left(D^*\right)=\frac{\delta}{1+\delta}\left[\alpha(2-\beta) v_L-v_H\right] s^{\prime}\left(D^*\right) .
\label{eqn:A4}
\end{equation}

Now, since $v_L>v_H / \alpha(2-\beta)$, it follows that $\alpha(2-\beta) v_L-v_H>0$, and thus, there is a positive value of $D^*$ that satisfies (\ref{eqn:A4}).
As before, comparing (2) and (A4), since $c^{\prime \prime}(D)>0, D^*<D^{* *}$ if

$$
\left[\alpha(2-\beta) v_L-v_H\right]<v_L,
$$
which simplifies to
$$
v_H>v_L(\alpha(2-\beta)-1),
$$
which is true for all $0<\alpha<1,0<\beta<1$ since $v_H>v_L$. The result follows. (Q.E.D.)

\textbf{\textit{Proof of Proposition 1}.} 
\begin{enumerate}[label=(\alph*)]
    \item By the envelope theorem, and the profit expressions in (\ref{eqn:A1}), it follows that in the third-party marketplace case:
$$
\frac{d \pi\left(D^*(\alpha)\right)}{d \alpha}=n_H \delta\left[(2-2 \beta) v_L s\left(D^{\prime}\right)\right]
$$

which is strictly positive for $\beta<1$. Also, rearranging (\ref{eqn:A2}) yields
\begin{equation}\tag{A5}
\frac{c^{\prime}\left(D^*\right)}{s^{\prime}\left(D^*\right)}=\frac{\delta}{1+\delta}\left[2 \alpha(1-\beta) v_L-v_H\right]
\label{eqn:A5}
\end{equation}
Since $c(D)$ is strictly convex and $s(D)$ is strictly concave, it follows that $\frac{c^{\prime}(D)}{s^{\prime}(D)}$ is strictly increasing in $D$ for all $D>0$. Notice that as $\alpha$ increases, the right-hand side of \ref{eqn:A5} is higher, thus requires a higher value of $\frac{c^{\prime}\left(D^*\right)}{s^{\prime}\left(D^*\right)}$ on the left-hand side, and consequently, a higher value of $D^*$.

Similarly, by the envelope theorem, and the profit expressions in (\ref{eqn:A1}), it follows that in the branded recommerce case:
$$
\frac{d \pi\left(D^*(\alpha)\right)}{d \alpha}=n_H \delta\left[(2-\beta) v_L s\left(D^{\prime}\right)\right],
$$
which is also strictly positive for $\beta \leq 1$. Correspondingly, rearranging (\ref{eqn:A4}) yields
\begin{equation}\tag{A6}
\frac{c^{\prime}\left(D^*\right)}{s^{\prime}\left(D^*\right)}=\frac{\delta}{1+\delta}\left[\alpha(2-\beta) v_L-v_H\right]
\label{eqn:A6}
\end{equation}
and based on the same logic as above, a higher value of  $\alpha$ implies a higher value of $D^*$. The result follows. 

\item  Inspecting (\ref{eqn:A5}) and (\ref{eqn:A6}) above establishes that a higher value of $\beta$ induces a lower value of $\mathrm{D}^*$, by the same argument involving $\frac{c^{\prime}(D)}{s^{\prime}(D)}$ being strictly increasing in D. Applying the envelope theorem to the profit expressions in (\ref{eqn:A1}) and (\ref{eqn:A3}) yields:
$$
\frac{d \pi\left(D^*(\beta)\right)}{d \beta}=-2 n_H \delta \alpha v_L s\left(D^{\prime}\right)<0
$$
in the case of the third-party marketplace, and 
\begin{equation}\tag{A7}
\frac{d \pi\left(D^*(\beta)\right)}{d \beta}=-n_H \delta \alpha v_L s\left(D^{\prime}\right)<0
\label{eqn:A7}
\end{equation}
\end{enumerate}
(Q.E.D.)\\
\textbf{\textit{Proof of Proposition 2}.} 
\begin{enumerate}[label=(\alph*)]
    \item Recall from (\ref{eqn:A5}) and (\ref{eqn:A6}), optimal durability $D^*$ satisfies
$$
\frac{c^{\prime}\left(D^*\right)}{s^{\prime}\left(D^*\right)}=\frac{\delta}{1+\delta}\left[2 \alpha(1-\beta) v_L-v_H\right],
$$
with a third-party marketplace and
$$
\frac{c^{\prime}\left(D^*\right)}{s^{\prime}\left(D^*\right)}=\frac{\delta}{1+\delta}\left[\alpha(2-\beta) v_L-v_H\right],
$$
with branded recommerce. Comparing the right-hand sides of (\ref{eqn:A5}) and (\ref{eqn:A6}), and using the fact that $2 \alpha(1-\beta)<\alpha(2-\beta)$ if $\alpha>0, \beta>0$, it follows that:
$$
\left[\alpha(2-\beta) v_L-v_H\right]>\left[2 \alpha(1-\beta) v_L-v_H\right] .
$$
Since $\frac{c^{\prime}(D)}{s^{\prime}(D)}$ is strictly increasing in $D$ as established in the proof of Proposition 1, the durability level that satisfies (\ref{eqn:A5}) must be higher than the durability level that satisfies (\ref{eqn:A6}). The result follows. 

\item Recall from (\ref{eqn:A7}) that by the envelope theorem,
$$
\frac{d \pi\left(D^*(\beta)\right)}{d \beta}=-n_H \delta \alpha v_L s\left(D^{\prime}\right)<0 .
$$
Under branded recommerce, the seller can choose $\beta$, and will thus choose the lowest feasible level $\beta=0$. The result follows.
\item  Compare the expressions for $\frac{d \pi\left(D^*(\alpha)\right)}{d \alpha}$ in (\ref{eqn:A5}) and (\ref{eqn:A6}). Since $\beta \geq 0$, it follows that

\begin{equation}\tag{A8}
n_H \delta\left[(2-\beta) v_L s\left(D^{\prime}\right) \geq n_H \delta\left[(2-2 \beta) v_L s\left(D^{\prime}\right)\right]\right.
\label{eqn:A8}
\end{equation}
However, part (b) establishes that under branded recommerce, $\beta=0$, and thus the inequality in (\ref{eqn:A8}) is strict for any third-party marketplace with a non-zero commission. Thus, the seller gets a larger increase in profits under branded recommerce from any fixed increase in $\alpha$ or equivalently, any fixed reduction in quality uncertainty, than with a third-party marketplace. The result follows. 

\end{enumerate}

\textbf{\textit{Proof of Lemma 5}.}

Note that (\ref{eqn:H_2}) implies that:
\begin{equation}
\left.\left.\max \left[v_{H}-p^{n}+1-\beta\right) p^{u}, v_{H} s(D)\right]=v_{H}-p^{n}+1-\beta\right) p^{u}.
\label{eqn:H_2_max}
\end{equation}

Rearranging (\ref{eqn:H_2}) yields 

\begin{equation}
[1-s(D)] v_H+(1-\beta) p^u \geq p^n
\label{eqn:H_2_rearrange}
\end{equation}

We can use (\ref{eqn:H_2_max}) to rewrite (\ref{eqn:H_1}) as

\begin{equation}
v_H-p^n+\delta\left[v_H-p^n+(1-\beta) p^u\right] \geq \alpha v_H s(D)-p^u
\label{eqn:H_1_rearrange_1}
\end{equation}

Rearranging (\ref{eqn:H_1_rearrange_1}) yields

\begin{equation}
\left[1-\frac{\alpha}{1+\delta} s(D)\right] v_H+\left(1-\frac{\delta}{1+\delta} \beta\right) p^u \geq p^n
\label{eqn:H_1_rearrange_2}
\end{equation}

Since $\frac{\alpha}{1+\delta}<1$ and $\frac{\delta}{1+\delta}<1$, if (\ref{eqn:H_2_rearrange}) is satisfied, (\ref{eqn:H_1_rearrange_2}) can be ignored. It follows that (\ref{eqn:H_2}) implies (\ref{eqn:H_1}).

From (\ref{eqn:L_12}), since there are more buyers than sellers in the pre-owned marketplace, it follows that

\begin{equation}
p^u=\alpha v_L s(D)
\label{eqn:p_u}
\end{equation}

(\ref{eqn:p_u}) and (\ref{eqn:H_2}) imply that the highest price the seller can charge for new goods is 

\begin{equation}
p^n=\alpha(1-\beta) v_L s(D)+v_H(1-s(D))
\label{eqn:p_n}
\end{equation}
Note that these are identical to the second-period prices obtained in the two-period model. 

We can use (\ref{eqn:p_u}) and (\ref{eqn:p_n}) to derive that

\begin{equation}
\max \left[v_L-p^n+(1-\beta) p^u, v_L s(D)\right]=v_L s(D)
\label{eqn:L_1_max}
\end{equation}

We can use (\ref{eqn:p_u}), (\ref{eqn:p_n}), (\ref{eqn:L_1_max}) together to simplify (\ref{eqn:L_1}) as:

\begin{equation}
\frac{v_L}{v_H}\leq\frac{1-s(D)}{1-[(1-\beta) \alpha-\delta] s(D)}
\label{eqn:L_1_frac}
\end{equation}

If (\ref{eqn:L_1_frac}) is true, we can establish that (\ref{eqn:L_1}) is satisfied and (\ref{eqn:L_2}) is thus implied. This completes the proof. (Q.E.D.)

\textbf{\textit{Proof of Proposition 3}.} 
Imagine the optimization problem faced by the seller when choosing their optimal durability level D. Let the first-period price be g1(D). Define 
 \begin{equation}
G_T(D)=\sum_{t=1}^{\infty} \delta^t n_H\left[\alpha(1-\beta) v_L s(D)+v_H(1-s(D))-c(D)\right]
\end{equation}
and
\begin{equation}
G_B(D)=\sum_{t=1}^{\infty} \delta^t n_H\left[\alpha v_L s(D)+v_H(1-s(D))-c(D)\right] .
\end{equation}
These are the steady-state profit streams as viewed from period 1 , under the third-party pre-owned marketplace $(\mathrm{T})$ and branded recommerce (B) respectively. The seller therefore chooses $\mathrm{D}$ to maximize $n_H g_1(D)+G_T(D)$ with a third-party marketplace, and $n_H g_1(D)+G_B(D)$ with branded recommerce. Under the parameter restrictions that ensure the seller will not want to shut out the type-L customers, $G_B{ }^{\prime}(D)>G_T{ }^{\prime}(D)$ and $G_B{ }^{\prime \prime}(D)<G_T{ }^{\prime \prime}(D)$, which is sufficient to establish that durability is higher under branded recommerce. Standard comparative statics methods establish the rest of the results. (Q.E.D.)

 \end{APPENDIX}
%
%


\bibliographystyle{informs2014} 
\bibliography{recommerce_refs} 


\end{document}